\documentclass[12pt]{iopart}

\usepackage{graphicx}
\usepackage{dcolumn}
\usepackage{bm}
\usepackage{color}
\usepackage[T1]{fontenc}
\usepackage{mathpazo}
\usepackage{color,soul}
\usepackage{subfigure}
\usepackage{indentfirst}
\usepackage{hyperref}
\makeatletter
\@ifundefined{equation*}{}{%
  \expandafter\let\csname equation*\endcsname\relax
  \expandafter\let\csname endequation*\endcsname\relax
}
\makeatother
\usepackage[version=4]{mhchem}

\usepackage[titletoc,title]{appendix}
\begin{document}

\title[Ji Yung Ahn et al.]{Plasma Decay of Nanosecond Pulsed Laser-Produced Ar and Ar-H$_{2}$O Sparks at Atmospheric Pressure}

\author{Ji Yung Ahn, Jianan Wang, Tasnim Akbar Faruquee, and Marien Simeni Simeni}

\address{Department of Mechanical Engineering, University of Minnesota, 111 Church St SE, Minneapolis, MN, 55455, USA}
\ead{msimenis@umn.edu}
\vspace{10pt}
\begin{indented}
\item[]November 2025
\end{indented}

\begin{abstract}
Time-resolved diagnostics were applied to investigate free-electron properties in nanosecond laser-produced discharges sustained in atmospheric pressure Ar and in Ar-3\%H$_2$O. The discharges were generated using 23 ns, 1064 nm laser pulses. Broadband plasma imaging and laser Thomson scattering were combined with optical emission spectroscopy, with particular emphasis on the Stark broadening of the H$_{\alpha}$ and H$_{\beta}$ lines. The plasma exhibited a bright emission that persists for up to 30-40 $\mu$s after breakdown. Plasma emission was then followed by a very weak glow emission that persisted for up to 19 ms after breakdown. Peak electron number density of \textasciitilde{2} $\times$ 10$^{17}$ cm$^{-3}$ and electron temperature of \textasciitilde{7} eV were measured. An excellent agreement between both techniques was obtained regarding absolute electron number densities. The inferred free-electron temporal decay dynamics are consistent with processes dominated by ambipolar expansion and two- and three-body electron-ion recombination. These results provide benchmark data for modeling nanosecond laser discharges and demonstrate the reliability of combining Thomson scattering with Stark broadening in atmospheric laser sparks.
\end{abstract}

%
%
%
%
%

\section{Introduction}

Nanosecond pulsed laser-produced plasmas (LPPs) have been extensively studied due to their broad range of applications, which depend strongly on both the laser parameters and the target medium~\cite{radziemski2020lasers}. A general distinction can be drawn between plasmas generated from solid/liquid targets and those sustained in gases. The former are widely exploited for material processing applications such as surface modification \cite{Li_etal_2006, dePablosMartin_Hoche_2017}, thin-film deposition \cite{Ashfold_etal_2004}, and micromachining via laser ablation \cite{Knowles2007}, whereas gaseous LPPs are relevant for trace-species detection \cite{Tognoni_Cristoforetti_2014}, ignition of combustible mixtures \cite{Kopecek_etal_2003, Dumitrache_etal_2017}, high-speed flow control \cite{Limbach_2015}, and propulsion \cite{Horisawa_etal_2005, Phipps_2018}. At relativistic intensities on solid targets ($\geq$ 10$^{18}$ W/cm$^{2}$), nanosecond pulses are also used in the generation of relativistic electron beams \cite{bailly2018guiding}, inertial confinement fusion \cite{Betti2016, Hohenberger2012}, and laboratory astrophysics \cite{Remington2000, Woolsey2004}, while in gases, high-intensity (10$^{9}$-10$^{11}$ W/cm$^{2}$) LPPs are central to the development of extreme ultraviolet (EUV) and soft x-ray light sources \cite{Fiedorowicz_etal_2000}.\\
Considerable effort has been devoted to understanding the spatiotemporal dynamics of laser sparks in gases. Early studies examined breakdown thresholds, plasma morphology, and emission properties in air and noble gases \cite{longenecker2003laser, glumac2005temporal, harilal2004spatial}. More recent works have revealed complex, non-equilibrium behavior of plasma kernels, strongly influenced by multimode laser structures and hydrodynamic effects such as vorticity generation \cite{alberti2019laser, alberti2022non, dumitrache2020gas}. Computational models further highlighted the role of non-equilibrium kinetics and transport in shaping the plasma evolution \cite{munafo2020computational, pokharel2023self, pokharel2025characteristics}.\\
Diagnostics of nanosecond LPPs have historically relied on optical emission spectroscopy (OES) because of its simplicity and sensitivity \cite{aragon2008characterization}. Stark broadening of hydrogen and argon lines has been used extensively to infer electron densities \cite{dzierzega2019experimental}, though uncertainties arise from line-shape modeling and plasma inhomogeneities. Complementary active techniques such as laser Thomson scattering (LTS) and Rayleigh scattering provide direct access to electron densities and temperatures \cite{dzierzega2010thomson, mendys2011investigations, zhang2019investigations}. However, LTS remains experimentally challenging due to its weak signal, which has limited its systematic application, especially at atmospheric pressure in argon.\\
Most recent diagnostic studies of laser filaments and optical discharges have focused on femtosecond laser sparks in air, motivated by applications in remote sensing and lightning control \cite{bak2024two, bak2025plasma}. However, comparatively fewer works have addressed nanosecond-pulsed LPPs in argon at atmospheric pressure, despite early demonstrations such as the experiments of Weyl and Rosen (1985) using the third harmonic of a Nd:YAG laser \cite{weyl1985laser}. Existing reports in argon emphasize plasma expansion and radiative lifetimes \cite{harilal2004spatial, mendys2011investigations}, but comprehensive time-resolved measurements of free-electron properties under controlled conditions remain limited.\\
In this work, we present time-resolved Thomson scattering and optical emission spectroscopy measurements of nanosecond laser-produced plasmas sustained in atmospheric-pressure argon and Ar–3\% H$_2$O. The diagnostics are complemented by broadband plasma imaging, enabling a combined assessment of electron densities, electron temperatures, and plasma emission dynamics. Particular attention is given to comparisons between electron densities inferred from TS and Stark broadening of H$_\alpha$ and H$_\beta$, thereby addressing long-standing questions regarding the reliability and consistency of these complementary techniques. The paper is organized as follows: Section 2 describes the experimental setup, Section 3 presents the results, discusses the findings in the context of scaling laws and previous literature. We summarize and conclude in Section 4.

\section{Experimental Set-up}
\begin{figure}
    \centering
    \includegraphics[width=0.95\textwidth]{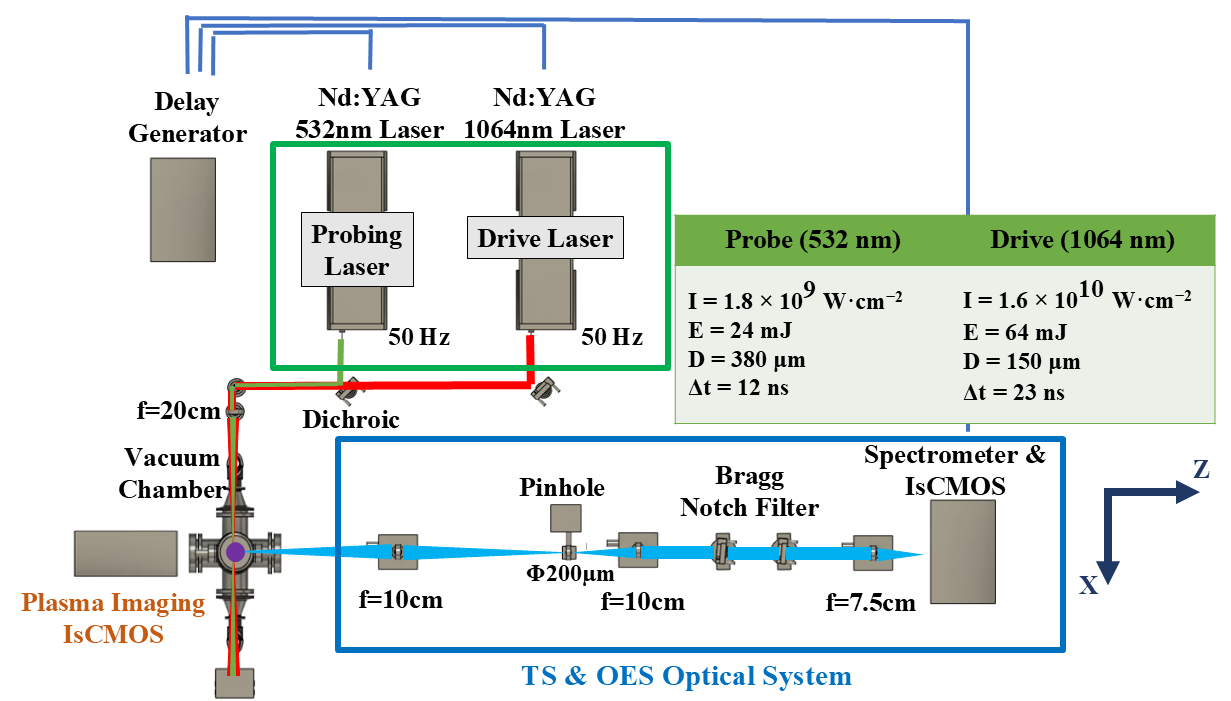}
    \caption{Schematic of the laser Thomson scattering experimental setup. Features of the drive and probe laser beams are also provided. OES measurements were performed using the same collection path as LTS. Imaging is performed on the opposite side to LTS measurements.}
    \label{TSetup}
\end{figure}

\begin{figure}
    \centering
    \includegraphics[width=0.75\textwidth]{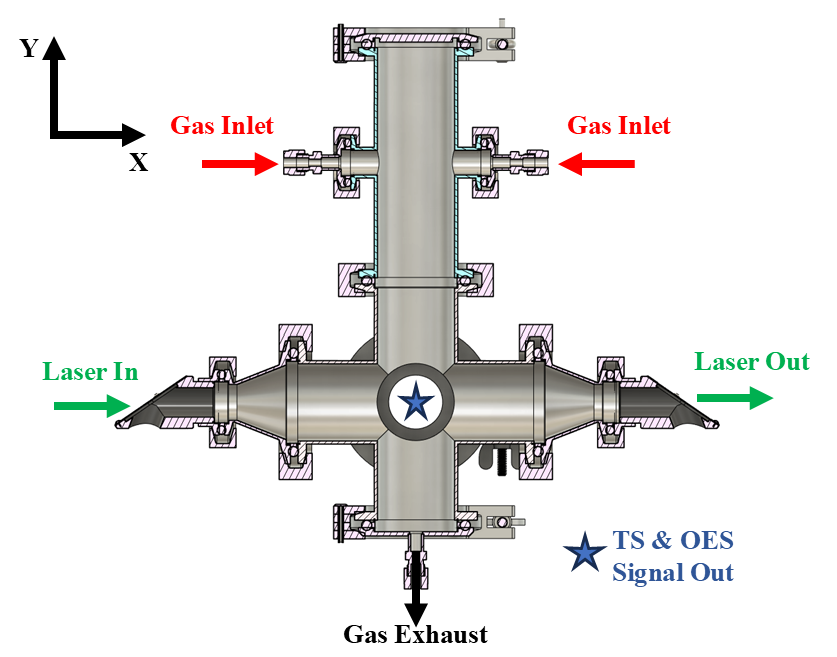}
    \caption{Schematic of the vacuum chamber hosting the laser sparks in argon. Laser scattering signals are collected through the viewport at the center while plasma imaging is conducted through the opposing viewport.}
    \label{Chamber}
\end{figure}

A schematic of the experimental arrangement is shown in Fig.~\ref{TSetup}. The central component is a custom-built test vessel (detailed in Fig.~\ref{Chamber}), consisting of an aluminum vacuum chamber with a KF50 six-way cross flange. The vessel is fitted with two opposing axial (\textit{x}-direction) Brewster windows (ThorLabs, BW2002) and two side-mounted Kodial glass viewports (LDS vacuum, NW50-200-ZVP). Quarter-inch gas lines, valves, and a vacuum pump, allow the chamber to be evacuated and filled with different high-purity gases. In this work, the vessel was operated at atmospheric pressure, containing either pure argon or a mixture of argon with 3\% water vapor. The latter mixture was prepared using a bubbler following the procedure in \cite{verreycken2012time, verreycken2013spectroscopic}. The volumetric flow rate of the argon gas was fixed at 1000 sccm.\\
\noindent Laser-produced plasmas (LPPs) were generated at the vessel center using the 1064 nm fundamental beam output of a 50 Hz nanosecond pulsed neodymium:yttrium aluminum garnet (Nd:YAG) laser (Continuum/Amplitude Laser, Powerlite 9050). The drive beam was horizontally polarized and focused to a \textasciitilde{150} $\mu$m diameter spot, with typical pulse energies of 64 mJ, corresponding to an intensity of approximately 1.6$\times$ 10$^{10}$ W/cm$^{2}$ at focus.\\
\noindent Thomson scattering was performed using the frequency-doubled (532 nm) output of a second, co-propagating 50 Hz nanosecond pulsed Nd:YAG laser (Big Sky/Quantel, Ultra-CFR/Emas). The probe beam was horizontally polarized, with \textasciitilde{24} mJ pulse energy and a \textasciitilde{380} $\mu$m focal spot, corresponding to a fluence of \textasciitilde{24} J/cm$^{2}$ and an intensity of \textasciitilde{1.8}$\times$ 10$^{9}$ W/cm$^{2}$. Drive and probe beam pulse energies were determined using a thermopile optical power-meter (Newport, 919P-030-18). Temporal profiles of both lasers were characterized using a 150 ps rise-time silicon photodetector (ThorLabs, DET025AL) coupled to a 2.5 GHz-bandwidth oscilloscope (Keysight, EXR258A). Focal spot sizes were measured using the traveling knife-edge technique \cite{de2009measurement}.\\
The drive and probe lasers were synchronized using an 8-channel digital delay generator (Stanford Research Systems, DG645), serving as the master clock for time-resolved measurements. Both beams were focused into the vessel by a $f$ = 20 cm achromatic plano-convex doublet lens (ThorLabs, AC254-200-AB, AR-coated) and were terminated in a beam trap after passing through the opposite Brewster window.

\noindent Probe light scattered at 90° was collected with a 1-inch diameter, $f$ = 10 cm achromatic plano-convex lens (ThorLabs, LA1509-A, AR-coated) in a $2f$–$2f$ configuration, forming a unity-magnification image of the scattering volume onto a 200 $\mu$m diamond pinhole (Fort Wayne Wire Die), which served as a spatial filter. Transmitted light was collimated with another $f$ = 10 cm 1-inch plano-convex lens (ThorLabs, LA1509-A, AR-coated) and spectrally filtered using two volume Bragg grating (VBG) notch filters (OptiGrate Corp., 11 mm, 532 nm BNF-OD-4) in series to suppress elastic scattering and stray 532 nm light. The filtered, collimated light beam was then focused onto the motorized entrance slit of a 32.8 cm focal length Czerny-Turner spectrometer (Andor Technology Inc, Kymera 328I) equipped with a 4-grating turret including ruled gratings of 150, 300, 600, and 1200 grooves/mm, all blazed at 500 nm. A time-gated, intensified sCMOS camera (Andor Technology Inc., iSTAR-sCMOS-18U-63, 2560x2160 pixels, 6.5 $\mu$m pixel size) was used for signal acquisition. The 1200 grooves/mm grating was used for both Thomson scattering and optical emission spectroscopy (OES), with the latter conducted through the same collection optics.\\
For spectral calibration and resolution measurements, low-pressure Hg-Ar (Oriel/Newport, 6035) and Ne (Oriel/Newport, 6032) lamps were employed. Time-resolved plasma images were obtained through the viewport opposite the Thomson scattering collection path, using the same intensified sCMOS camera fitted with a microfocus lens (Nikon, Micro-Nikkor 105 mm, 400-700 nm wavelength range). Finally, preliminary time-averaged OES measurements were conducted through the latter viewport using a miniature, lower spectral resolution pocket spectrometer (Ocean Insight, FLAME-S-UV-VIS-ES) covering the 200-850 nm spectral range.
 
\section{Results and Discussions}

\subsection{Broadband Plasma Imaging}
\begin{figure}
    \centering
    \includegraphics[width=1.00\textwidth]{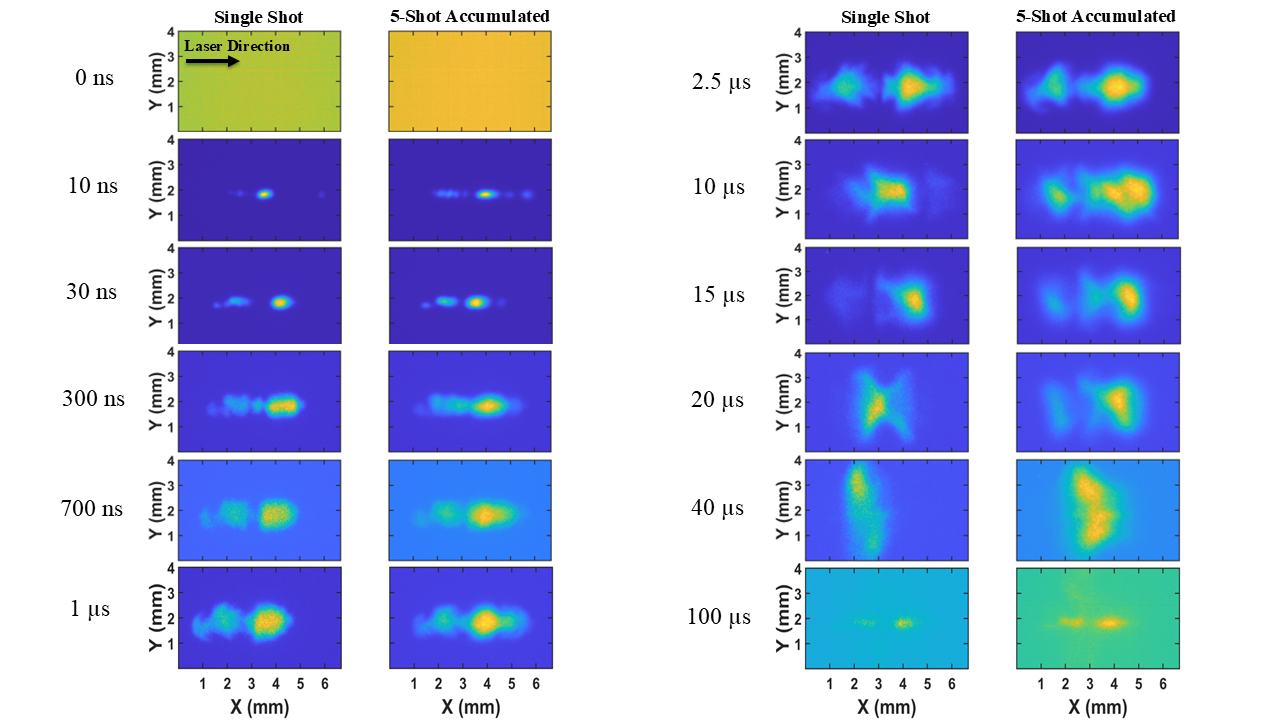}
    \caption{Time-resolved plasma images in pure Ar. The camera gate width is 5 ns. Single-shot and 5-shot accumulated plasma images are shown. Time zero is taken as the arrival time of the 1064 nm drive beam at the focal point inside the test vessel.}
    \label{ColDry}
\end{figure}

The temporal dynamics of the optical discharges sustained in pure Ar is depicted in Fig.~\ref{ColDry}. Single-shot and 5-shot averaged plasma images recorded using 5 ns camera gate widths are compared for time delays spanning from the first observed light up to 100 $\mu$s after the arrival of the drive beam at the center of the chamber. The latter range of time delays was chosen to overlap with the period during which Thomson scattering measurements were performed (from 500 ns to 20 $\mu$s). Clearly, plasma images in Fig.~\ref{ColDry}, when combined with those shown in Appendix A, reveal long-lasting light emission up to 19 ms after the inception of the optical discharge. This would initially come as a surprise, especially given that the 1064 nm drive laser beam responsible for the generation of the plasma was only about 20 ns in duration. Indeed, such behavior contrasts starkly with that of nanosecond pulsed electrical discharges sustained in atmospheric pressure argon or helium with a comparable amount of energy deposited in the gas \cite{wu2021spatially, van2025electron}. It can be hypothesized that these results arise from the fact that laser sparks typically produce higher peak electron densities, stronger hydrodynamic heating, and a long-lived metastable-driven afterglow that prolongs the emission~\cite{radziemski2020lasers}. Longenecker \textit{et al}, for instance, observed emission up to 4 ms after the initiation of the optical discharge. In their experiments, they also employed the fundamental of a nanosecond pulsed Nd:YAG laser to generate a plasma in a vessel containing argon (or Ar-N$_2$ mixtures) at atmospheric pressure~\cite{longenecker2003laser}. Although the intensity of their drive laser (10$^{11}$ W/cm$^2$) was about one order of magnitude higher than ours, their drive laser repetition rate was only 10 Hz (compared to 50 Hz for our experiments). Despite the differences observed with electrical discharges, such long-lasting behavior is commonly observed in laser-produced plasmas over a vast parameter space range, which includes LPPs sustained at high and low pressures as well as LPPs generated at different intensities \cite{weyl1985laser, longenecker2003laser, mendys2011investigations}.\\
At \textit{t} = 10, 30 ns, the discharge appears to be elongated in the laser propagation direction (\textit{x}-direction). It is constituted by what appear to be 2-3 discontinuous bright spots (plasma kernels). When single-shot and accumulated plasma images are compared, it is clear that the discharge is not perfectly reproducible; this is probably due to the stochastic nature of the optical breakdown process. Indeed, the statistical nature of the seed-electron availability at such a relatively low laser pulse repetition rate (50 Hz i.e. 20 ms time periods) combined with beam-focusing instabilities (laser-pointing jitter and energy fluctuations) and possibly the presence of trace contaminants~\cite{pasquini2007laser, fu2019investigation} could explain the discharge occurrence at slightly different locations along the laser beam from shot to shot. For our experiments, we measured a time jitter of about 2 ns for the drive laser beam, while a pulse energy fluctuation of about 14\% was observed. Note that although both single-shot and 5-shot accumulated plasma images seem to point to a discontinuous plasma, with separated lobes, this is in fact a result of the image-processing procedure. The discharge obviously displays strong gradients in emitted intensity along the \textit{x}-direction, but plasmas also exist between the observed lobes.\\
Later, from \textit{t} = 300 ns to 1 $\mu$s, the discharge grows slightly in the \textit{x}-direction (both forward and backward), while growing substantially in the \textit{y}-direction from \textasciitilde{0.3} to \textasciitilde{1} mm. Next, from \textit{t} = 2.5 to 15 $\mu$s, the discharge appears to exhibit "branches" in the \textit{y}-direction while maintaining the elongated feature in the \textit{x}-direction. It should be noted that a maximum axial length \textasciitilde{5} mm was reached around \textit{t} = 10 $\mu$s. At \textit{t} = 20, 40 $\mu$s, the discharge appears to have stopped propagating axially. Interestingly, at \textit{ t} = 40 $\mu$s it even showcases a predominant \textit{y}-direction orientation, reaching \textasciitilde{4} mm in the \textit{y}-direction for only \textasciitilde{1} mm in the \textit{x}-direction. Finally, at \textit{t}= 100 $\mu$s, light emitted by the discharge is very dim and the discharge appears to have returned to a propagation in the axial direction alone. Looking at Appendix A, light emission pattern remains nearly unchanged from \textit{t} = 100 $\mu$s to \textit{t} = 19 ms but completely ceases before the next laser pulse (see plasma image at \textit{t} = 0 ns). It is hypothesized that at the aforementioned times, emission traces neutral gas heating rather than plasma. A higher camera gain setting was applied to capture plasma images at \textit{t} $\geq$ 100 $\mu$s.\\ 
Multiple previous experimental studies have reported the formation of two-lobe structures for LPPs generated in high-pressure gases~\cite{bindhu2003laser, harilal2004spatial, glumac2005temporal, mendys2011investigations, pokrzywka2012laser}. These studies generally agree that the observed two lobes are counter-propagating. The so-called "rear lobe" grows towards the drive laser beam (negative \textit{x}-direction), and the "front lobe" propagates in the same direction as the drive laser beam (positive \textit{x}-direction). However, there is no consensus on the mechanism responsible for the formation of these two-lobe structures. While a series of papers by Tsuda and Yamada argue in favor of plasma lensing effects arising from spatial gradients of the plasma refractive index~\cite{tsuda1997observation, tsuda2000mechanism}, Munafò and co-authors suggested a different mechanism. They proposed that the two-lobed plasma kernel arises from multiphoton ionization-triggered and free-free electron-ion inverse Bremsstrahlung-sustained heating that generates two hydrodynamic waves (forward/backward), whose nonlinear interaction produces and maintains the two-lobe structure without the need for plasma lensing effects~\cite{munafo2020computational}. We argue that both mechanisms may coexist. At sub-$\mu$s timescales, plasma refractive index gradients can lens the beam, while at later times, hydrodynamic shock interactions may dominate.\\
Interestingly, Nishihara \textit{et al} reported experimental observations of multi-lobe structures with the number of lobes greater than 2 when using a multimode drive laser beam (at 532 nm) in room air~\cite{nishihara2018influence}. When comparing their experimental and modeling results, they attributed these observations to the growth of local microstructures following fluctuations in the breakdown location triggered by the 500-800 MHz mode beating of the drive laser. Although our experimental observations of global modulations of the plasma envelope agree well with those of Nishihara \textit{et al}, the temporal profile of our laser did not exhibit mode beating (see Appendix B). Interestingly, our drive laser spatial profile (also shown in Appendix B) revealed a flat top beam profile with inhomogeneities. We therefore hypothesize that these local microstructures arise from the 1064 nm  drive laser spatial beam profile, rather than its temporal profile.

\begin{figure}
    \centering
    \includegraphics[width=1.00\textwidth]{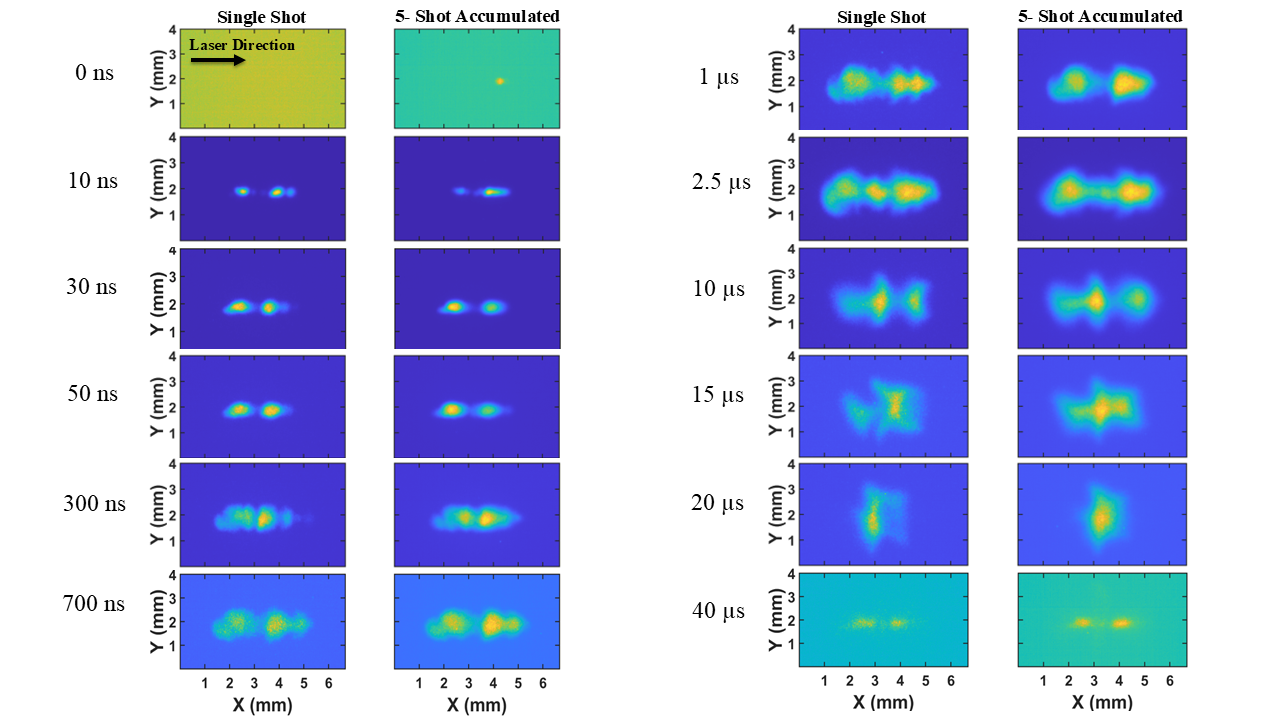}
    \caption{Time-resolved plasma images in Ar-H$_{2}$O. The camera gate width is 5 ns. Single-shot and 5-shot accumulated plasma images are shown. Time zero is taken as the arrival time of the 1064 nm drive beam at the focal point inside the test vessel.}
    \label{ColWet}
\end{figure}

\noindent Time-resolved plasma images in Ar-H$_{2}$O are shown in Fig~\ref{ColWet}. 
When single-shot and 5-shot averaged plasma images are compared, it is readily observed that the discharges generated in Ar-3\%H$_2$O are more reproducible from shot to shot than the discharges generated in pure Ar. This could be related to enhanced absorption of laser energy in the focal volume, in the presence of a larger concentration of water vapor molecules (since Ar is mostly transparent at 1064 nm), together with the fact that water molecules exhibit lower effective photoionization thresholds~\cite{liu2023influences}. In fact, the 12.6 eV ionization energy of water vapor is about 3 eV lower than that of argon (15.76 eV). This leads to an order of magnitude lower breakdown threshold in water vapor compared to that of argon, at atmospheric pressure. In general, strong similarities can be observed between single-shot and 5-shot averaged plasma images up to \textit{t} = 40 $\mu$s. Similarly to the pure argon case, the spatial structure of the emitted light in Ar-H$_2$O remains practically unchanged from \textit{t} = 40 $\mu$s to \textit{t} = 19 ms (see Appendix A). Still, a very similar temporal dynamics to that in the pure Ar case is observed for the Ar-H$_{2}$O case. A noticeable difference from the case of pure Ar is that the plasma contracts and cools faster in Ar-H$_2$O. For example, while the return to the sole axial propagation of the plasma occurs at \textit{t} = 100 $\mu$s in pure Ar, it occurs at \textit{t} = 40 $\mu$s in Ar-H$_2$O. This indicates earlier quenching of the plasma in Ar-H$_2$O compared to the pure Ar case. Early quenching of the plasma in the presence of water vapor molecules has been previously observed comparing electrical discharges in He and He-H$_{2}$O \cite{simeni2016electron}.

\subsection{Rotational Raman and Laser Thomson Scattering}
\begin{figure}
    \centering
    \includegraphics[width=0.65\textwidth]{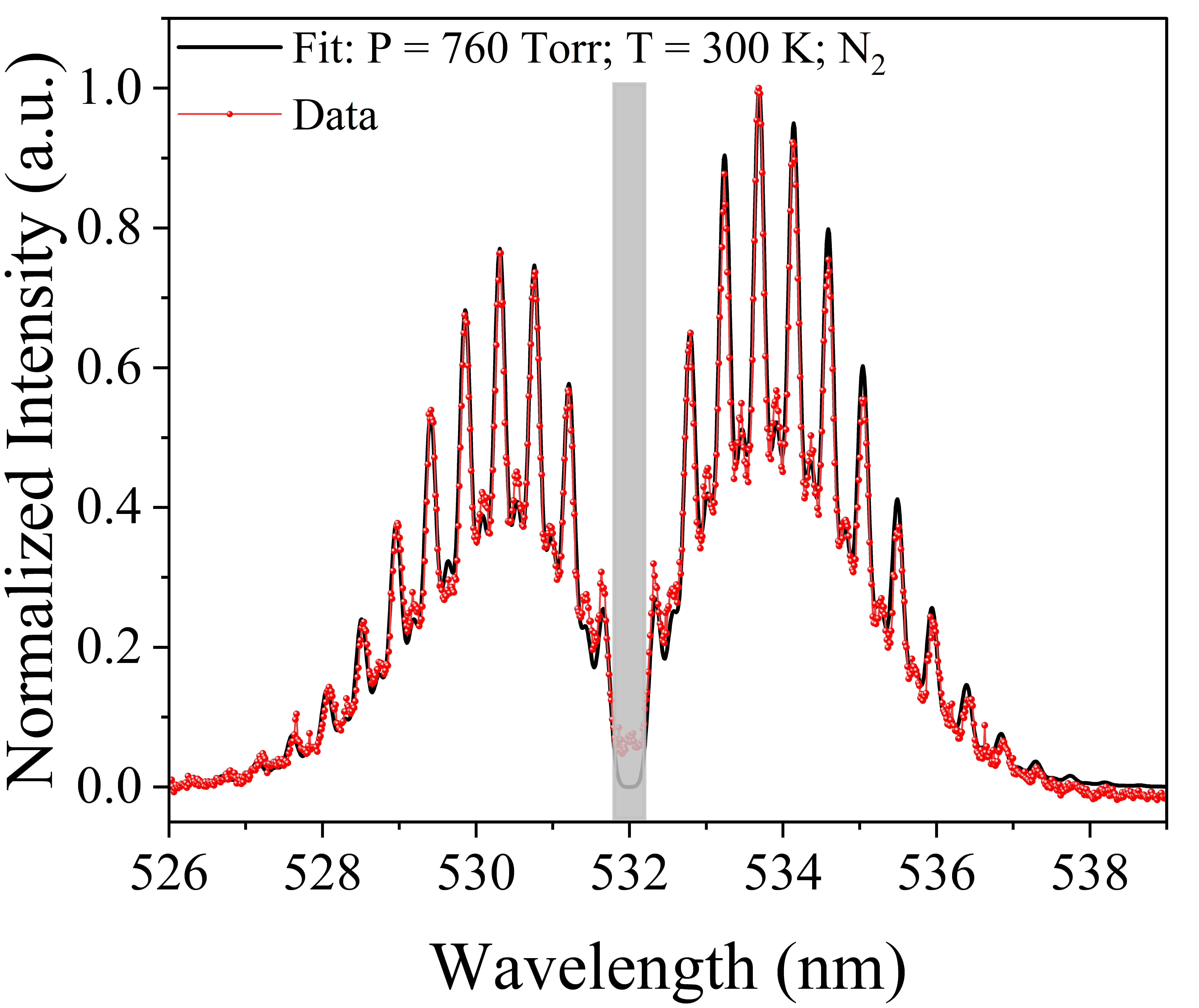}
    \caption{Sample pure rotational Raman spectrum in N$_{2}$ at 760 Torr and room temperature. The camera gate width is 20 ns. Spectrometer entrance slit width = 50 $\mu$m. The spectral resolution is about 190 pm. 30,000 laser shots were accumulated, with 30 seconds of exposure time repeated 20 times. The shaded region corresponds to the part of the spectrum rejected by the notch filter.}
    \label{N2}
\end{figure}

As an initial test of the efficiency of the laser scattering collection system, the test vessel was filled with pure N$_2$ at 760 Torr and room temperature. A pure rotational Raman spectrum featuring a 190 pm spectral resolution was recorded using the 1200 grooves/mm grating of the quad turret and a 50 $\mu$m spectrometer entrance slit width. The spectrum displayed in Figure~\ref{N2} was accumulated on-chip over 30,000 laser shots with a camera gate width of 20 ns. In greater detail, a 30 second exposure time (1,500 laser shots) was employed, and this was repeated 20 times. It can be readily observed that Rayleigh scattering and stray light at 532 nm, were effectively mitigated by the VBG notch filter strategy \cite{vincent2018compact}. A synthetic spectral fit built following the procedure highlighted in \cite{van2012laser} is overlaid on top of the experimental data points. The adequacy of the laser scattering setup is supported by the excellent agreement between experimental data and synthetic fit, with the measured spectrum preserving the expected intensity ratio between Stokes and Anti-Stokes lines at room temperature. N$_2$ rotational Raman will be later used for absolute calibration of the LTS results when in the non-collective scattering regime (see further).

\begin{figure}
    \centering
    \includegraphics[width=1.00\textwidth]{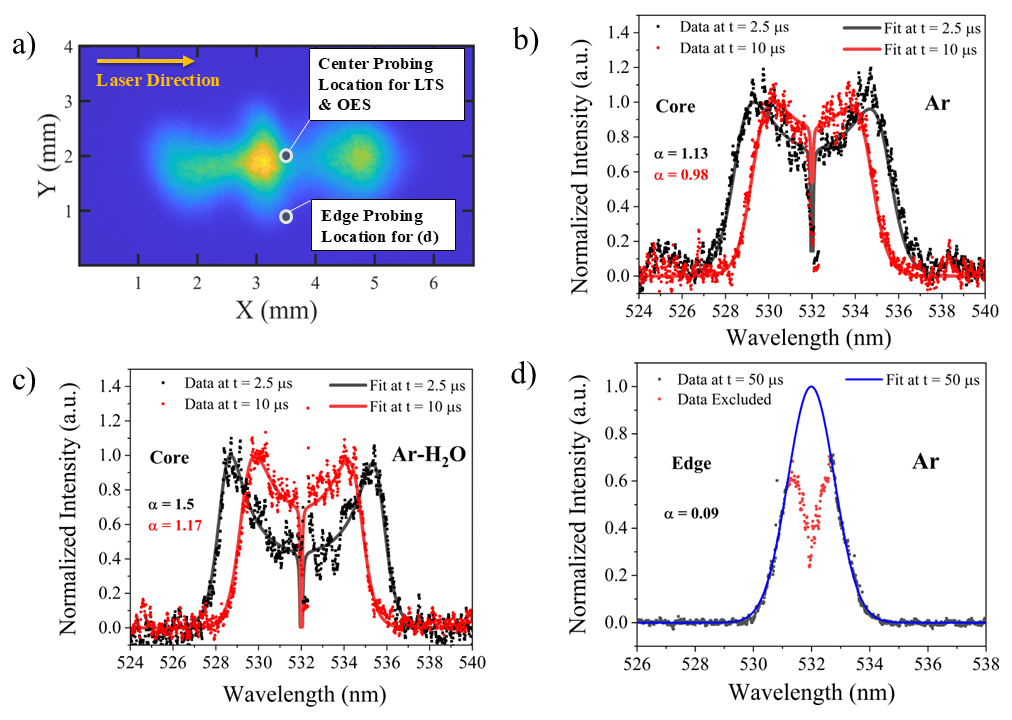}
    
    \caption{a) Measurement location for LTS measurements. b) Sample LTS spectra at \textit{t} = 2.5, 10 $\mu$s in Ar. c) Sample LTS spectra a \textit{t} = 2.5, 10 $\mu$s in pure Ar-H$_2$O.  d) Sample LTS spectra a \textit{t} = 50 $\mu$s in pure Ar at edge location. 30,000 laser shots were accumulated for each spectrum. The camera gate width is 20 ns. Spectrometer entrance slit width = 360 $\mu$m. The spectral resolution is 270 pm.}
    \label{TS-1}
\end{figure}

\noindent Figure~\ref{TS-1}-(a) depicts the measurement location for the so-called "plasma core/center" and "plasma edge" laser-scattering experiments conducted in this work. The plasma core measurement location is indicated by the label box superimposed on top of the 5-shot accumulated plasma image taken in Ar-H$_2$O at \textit{t} = 10 $\mu$s. The plasma edge measurement location is located about 1 mm below that of the core location in the \textit{y}-direction. All of the time-resolved LTS measurements within this manuscript pertain to the plasma core location. Only one sample LTS spectrum collected at the plasma edge will be discussed.\\ 
The Thomson scattering probe volume corresponds to the intersection between the 1:1 projection of the pinhole onto the probe beam scattering plane and the probe beam itself. For our experiments, because the probe beam spot (380 $\mu$m) is almost twice as large as the size of the pinhole (200 $\mu$m), our LTS probe volume is completely defined by the size of the pinhole. 
For Thomson scattering measurements performed on such gaseous targets under the given drive laser beam irradiance and atmospheric pressure conditions, it is expected that the LTS signal collected at 90° scattering angle transitions from a collective scattering regime, at early time delays to a non-collective scattering regime at later times \cite{miles2020time}. This was indeed observed, based on the sample Thomson scattering spectra provided in Figures~\ref{TS-1}-(b), (c), (d), respectively. The degree of "collectiveness" of these spectra is indicated by the so-called scattering parameter (or the Salpeter parameter):
\begin{equation}
    \alpha = \frac{1}{k \lambda_{D}} \approx \frac{1}{4\pi \sin{(\theta/2)}}\frac{\lambda_{i}}{\lambda_{D}}
\end{equation}
Where $k$, $\lambda_{D} \approx 7.4 \times 10^{-6} \sqrt{\frac{T_{e} [eV]}{n_e[10^{16} cm^{-3}]}}$, $\theta$, $\lambda_{i}$, $T_e$, $n_e$ denote the magnitude of the vector difference between the incident laser wavevector and the wavevector along the observation path, the Debye wavelength, the scattering angle, the probe laser wavelength, electron temperature and electron number density, respectively.\\
For $\alpha > 1$, scattering is in the collective regime while for $\alpha < 1$, scattering falls within the non-collective regime.  At 90° scattering angle and for a 532 nm probe beam, we therefore get the approximation:
\begin{equation}
\alpha \approx 8.1\times 10^{-1}\sqrt{\frac{n_e[10^{16} cm^{-3}]}{T_e[eV]}}
\end{equation}
Figure~\ref{TS-1}-(b) displays sample pure Ar LTS spectra in the collective and non-collective regimes ($\alpha \approx 1.13; 0.98$), which were measured at the plasma core location at \textit{t} = 2.5 and 10 $\mu$s, respectively. The spectra were collected using a 360 $\mu$m spectrometer entrance slit, resulting in a 270 pm spectral resolution and accumulated on-chip for 30,000 laser shots each (30 seconds exposure time repeated 20 times). Moreover, for each spectrum, the plasma emission background was collected separately over the same number of laser shots and subtracted. This subtraction approach only works if the plasma emission is reproducible enough over a large number of averages. In the collective regime, $n_e$ can be directly inferred from the spectral location of the so-called "electron plasma wave" (EPW) features (peaks), which are equidistant from the laser scattering central wavelength at 532 nm. Interestingly, in the collective regime, $T_e$ is inferred from the width of the EPW feature \cite{rocco2022applying, sheffield2010plasma}. In this work, rather than solely relying on the peak positions, we fit each of the complete measured spectrum using the procedure highlighted by Davies \cite{davies2019picosecond}. In pure Ar, at \textit{t} = 2.5 $\mu$s, we obtained electron number density and temperature values of 4.5 $\times$ 10$^{16}$ cm$^{-3}$ and 2.25 eV, respectively. At \textit{t} = 10 $\mu$s, we obtained electron number density and temperature values of 2.25 $\times$ 10$^{16}$ cm$^{-3}$ and 1.5 eV, respectively. Between these two moments in time, the plasma is hence cooling and recombining.\\
Similarly, for the Ar-H$_2$O case, Figure~\ref{TS-1}-(c) pictures sample LTS spectra in the collective regime ($\alpha \approx 1.5; 1.17$), which were measured at the plasma core location at \textit{t} = 2.5 and 10 $\mu$s, respectively. Here again, free-electrons of the plasma cool down while decaying. The inferred electron number densities were $7 \times 10^{16}$ cm$^{-3}$ and $2.9 \times 10^{16}$ cm$^{-3}$ at \textit{t} = 2.5 $\mu$s and 10 $\mu$s, respectively. Meanwhile, electron temperatures of 2 and 1.35 eV were obtained.\\
At \textit{t} = 50 $\mu$s, at the "plasma edge", the measured spectrum in pure Ar is in the non-collective regime ($\alpha \approx 0.1$). It can be fitted with a Gaussian function (see Figure~\ref{TS-1}-(d)). In this regime, we implemented the standard approach \cite{van2012laser} to retrieve plasma parameters from the fitted spectra: while the electron temperature is inferred from the Full Width at Half Maximum (FWHM) of the Gaussian fit, the electron number density is obtained based on the area under the curve of the Gaussian fit, calibrated in absolute using N$_2$ pure rotational Raman scattering (at atmospheric pressure and with the same entrance slit width as for LTS measurements). We inferred electron number density and temperature values of 1.6 $\times$ 10$^{14}$ cm$^{-3}$ and 1.2 eV, respectively.\\
It is important to point out that one should not expect to observe overlapping contributions of the LTS and water vapor pure rotational Raman signals. In fact, beyond the fact that high dissociation of molecular species is expected to occur during the timescales of interest of the present study, previous work has demonstrated that pure rotational Raman signals from water vapor are in general too weak to be observed during LTS experiments~\cite{simeni2016electron, yue2022electron}. The pure rotational Raman cross-sections for water vapor are \textasciitilde{15–20} times smaller than those of \ce{N2}~\cite{penney1974absolute, penney1976raman}.

\begin{figure}

    \includegraphics[width=1.00\textwidth]{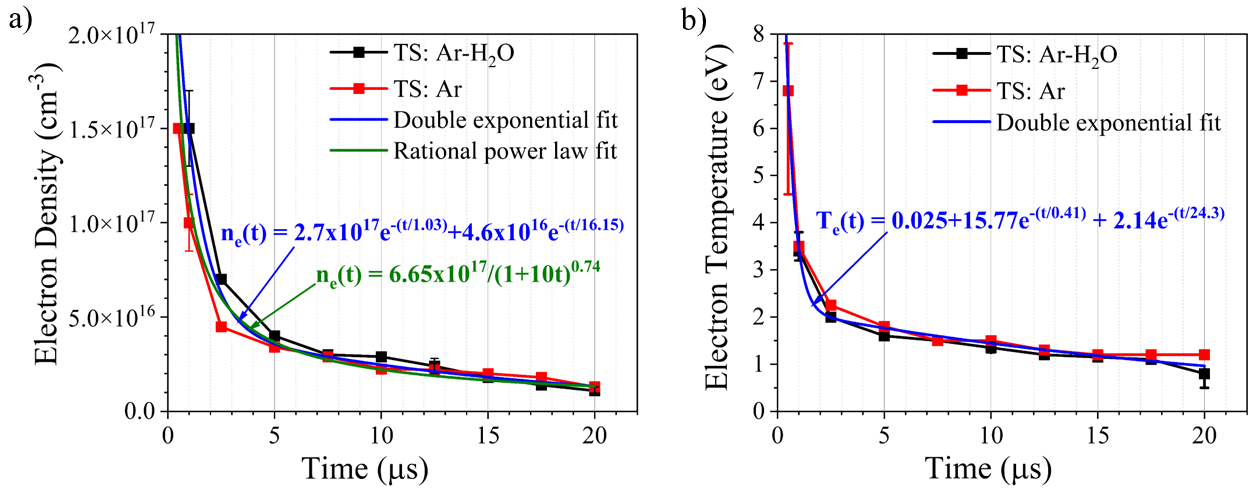}
    \caption{a) Time-resolved electron density inferred from Thomson scattering measurements in Ar and Ar-H$_2$O. b) Time-resolved electron temperature inferred from Thomson scattering measurements in Ar and Ar-H$_2$O. 30,000 laser shots were accumulated for each data point. The camera gate width is 20 ns. Spectrometer entrance slit width = 360 $\mu$m.}
    \label{ne-Te}
\end{figure}

\noindent Time-resolved $n_e$ and $T_e$ results inferred from LTS measurements in Ar and Ar-H$_2$O LPPs at the plasma core location are shown in Figure~\ref{ne-Te}-(a) and Figure~\ref{ne-Te}-(b), respectively. The measurements are reported for 500 ns $\leq$ \textit{t} $\leq$ 20 $\mu$s. At times earlier than 500 ns, the Thomson scattering signal is buried under the prevalent broadband continuum plasma emission (see further). From Figure~\ref{ne-Te}-(a), it appears that the electron number density values inferred from LTS are comparable for the Ar and Ar-H$_2$O cases, especially at \textit{t} $\geq$ 5 $\mu$s. For 500 ns $\leq$ \textit{t} $<$ 5 $\mu$s, electron number densities in Ar-H$_2$O are greater than those in pure Ar by up to 40-50 \%. This is attributed to the aforementioned enhanced absorption of the drive laser pulse energy in presence of water vapor molecules. In general, a decrease in $n_e$ is observed by approximately an order of magnitude from \textasciitilde{1.5} $\times$ 10$^{17}$ cm$^{-3}$ at \textit{t} = 500 ns to \textasciitilde{10}$^{16}$ cm$^{-3}$ at \textit{t}= 20 $\mu$s. This decay in the electron number density can be accurately fitted using a double-exponential decay function assuming the asymptotic condition that $n_e \rightarrow 0$ when $t \rightarrow +\infty$ (i.e. 20 ms in this case). This is justified by the lack of memory effects from a laser pulse to the following one, as evidenced by the full plasma images in Appendix A: 
\begin{equation}
    n_e (t) \approx 2.7 \times 10^{17} e^{-\frac{t}{1.03}} + 4.6 \times 10^{16} e^{-\frac{t}{16.15}},
\end{equation}
Where $n_e$ and $t$ are expressed in cm$^{-3}$ and $\mu$s units, respectively. This functional fit was added to Figure~\ref{ne-Te}-(a).\\
Two distinct characteristic times emerge from the fit: $\tau_{n1}$ $\approx$ 1 $\mu$s, characterizing the initial fast decay in the electron number density, and then $\tau_{n2}$ $\approx$ 16 $\mu$s characterizing the following slower decay.\\ 
Although it facilitated timescale analysis, it is important to note that the double-exponential decay function was chosen empirically. From a purely kinetic point of view, a rational power-law form~\cite{dogariu2013versatile} should be employed instead. Rational power-law fitting was carried out and an excellent agreement was achieved using the following functional form:

\begin{equation}
    n_e (t) \approx \frac{6.65 \times 10^{17}}{(1+10t)^{0.74}}
\end{equation}
Where $n_e$ and $t$ are expressed in cm$^{-3}$ and $\mu$s units, respectively. This functional fit was also added to Figure~\ref{ne-Te}-(a). The fitted exponent of $0.74$ (not a pure $\frac{1}{2}$ or $1$) suggests that electron loss is dominated by competing first order and second order processes. We can list for example: (1) three-body electron-ion recombination (3BR), (2) two-body electron-ion radiative recombination (2BR), (3) diffusion/ambipolar expansion, and (4) dissociative recombination in presence of water molecules.\\
In more detail, the three-body electron-ion recombination reaction can be expressed as:

\begin{equation}
\ce{e- + e- +Ar+ -> Ar + e-}
\end{equation}
Intuitively, this electron loss mechanism would be expected to dominate early in time, at high $n_e$ values since $\frac{dn_e}{dt} \propto n_e^2$. However, the latter analysis does not account for the electron temperature dependence of the reaction rate coefficient~\cite{bogaerts2003modeling}: 

\begin{equation}
k_{3BR} = 10^{-19}\left(\frac{T_e}{300}\right)^{-4.5}
\end{equation}
Where $T_e$ is in Kelvin units and $k_{3BR}$ has cm$^6$/s units. Therefore, it is clear that 3BR is not favored at high $T_e$ values, which is important since at high values of $n_e$, $T_e$ is also high. Assuming quasi-electroneutrality of the plasma ($n_e \approx n_i$), where $n_i$ denotes the number density of Ar$^+$ ions, we can estimate a characteristic time for 3BR: 

\begin{equation}
\tau_{3BR} \approx \frac{1}{k_{3BR} n_e^2} \approx 10^{19}\left(\frac{T_e}{300}\right)^{4.5}\frac{1}{n_e^2}
\end{equation}
For $n_e = 1.5 \times 10^{17}$ cm$^{-3}$ and $T_e$ = 6 eV, we get $\tau_{3BR} \approx 20$ $\mu$s. In contrast, for $n_e = 5 \times 10^{16}$ cm$^{-3}$ and $T_e$ = 2 eV, we get $\tau_{3BR} \approx 1.3$ $\mu$s. It is therefore evident that while 3BR does not significantly contribute to the initial fast decay in electron number density, it is expected to definitely play a role in the longer timescale decay.\\
The two-body radiative recombination reaction between free-electrons and Ar$^+$ ions can be written as:

\begin{equation}
\ce{e- + Ar+ -> Ar + h\nu}
\end{equation}
Where $h$ is Planck's constant and $\nu$ is the frequency of the emitted photon. Assuming the two-body radiative recombination rate coefficient ($\alpha_{2BR}$) is known, a characteristic time can be inferred as:

\begin{equation}
   \tau_{2BR} \approx \frac{1}{\alpha_{2BR} n_e^2}
\end{equation}
Using fit coefficients provided by Badnell~\cite{badnell2006radiative}, for the functional form of the electron temperature dependence of $\alpha_{2BR}$ proposed by Verner and Ferland~\cite{verner1995atomic}, we obtained $\alpha_{2BR} (T_e= 4 eV) \approx 4.1 \times 10^{-11}$ cm$^3$/s. This value increases to about 6.5 $\times 10^{-11}$ cm$^3$/s at $T_e = $ 2 eV and decreases to $\approx$ 3.1 $\times 10^{-11}$ cm$^3$/s at $T_e = $ 6 eV. Consequently, for $n_e = 1.5 \times 10^{17}$ cm$^{-3}$ and $T_e$ = 6 eV, we get $\tau_{2BR} \approx 0.2$ $\mu$s. In contrast, for $n_e = 5 \times 10^{16}$ cm$^{-3}$ and $T_e$ = 2 eV, we get $\tau_{2BR} \approx 0.3$ $\mu$s. Therefore, 2BR is much faster than 3BR under our experimental conditions. Moreover, 2BR seemingly contributes to the observed fast initial decay in free electron number densities.\\
Now, let us turn our attention to estimating the timescale of free electron number density decay as a result of ambipolar/hydrodynamic expansion of the plasma volume:

\begin{equation}
    \tau_{exp} \approx \frac{L}{c_s} \approx \frac{L}{\sqrt{kT_i/m_i}}
\end{equation}
Where $c_s$, $T_i$, $m_i$, $L$ are the ion-acoustic speed, ion temperature, ion mass and plasma characteristic length, respectively. If we assume thermal equilibrium at \textit{t} = 500 ns, we can write $T_i = T_e$. Therefore, for $T_i =$ 6 eV, $c_s$ $\approx 3.8 \times 10^3$ m/s. For characteristic axial lengths of $2-5$ mm, we obtain $\tau_{exp} \approx 0.52-1.3$ $\mu$s. This timescale is consistent with that of the initial fast decay.\\
In the presence of water vapor molecules, it is also expected that dissociative recombination could play a role: $\ce{e- + H2O+ -> OH + H}$ or $\ce{e- + H2O+ -> H2 + O}$. However, this mechanism is likely negligible for our experimental conditions since the experimentally measured timescales of electron number density decay in pure Ar and in Ar-\ce{H2O} cases do not show noticeable differences.\\
Taken together, the analysis above suggests that while the initial fast decay in electron number density is dominated by ambipolar/hydrodynamic expansion of the plasma together with two-body radiative recombination, the slower decay is affected by both of the aforementioned mechanisms, with the additional contribution of three-body recombination.\\
The temporal dynamics of the electron temperature is shown in Figure~\ref{ne-Te}-(b). Clearly, an almost perfect overlap between results in pure Ar and Ar-\ce{H2O} is observed. Similarly to $n_e$, a two-step decrease in $T_e$ was also observed. First, a very rapid decrease from \textasciitilde{7} eV at \textit{t} = 500 ns to \textasciitilde{2} eV at \textit{t} = 2.5 $\mu$s. This very fast decrease is followed by a slower decrease in electron temperature from \textasciitilde{2} eV at \textit{t} = 2.5 $\mu$s to \textasciitilde{1} eV at \textit{t} = 20 $\mu$s. Here again, an excellent double-exponential decay fit was obtained and a top layer was overlaid on the experimental data in Figure~\ref{ne-Te}-(b). The double-exponential fit for the temperature decay was obtained by assuming that electrons completely cool down to room temperature before the advent of the next drive laser pulse ($T_{e} \rightarrow 0.025$ eV (300 K) when $t \rightarrow +\infty$): 
\begin{equation}
T_e(t) \approx 0.025+ 15.77e^{-\frac{t}{0.41}} + 2.14 e^{-\frac{t}{24.3}},
\end{equation}
 Where $T_e$ and $t$ are expressed in eV and $\mu$s units, respectively. In this case, the characteristic times are $\tau_{T1}$ $\approx$ 0.4 $\mu$s and $\tau_{T2}$ $\approx$ 24 $\mu$s, respectively. Multiple previous studies have reported the agreement of multi-exponential fits with experimental data pertaining to electron temperature~\cite{mendys2011investigations} or gas temperature measurements~\cite{longenecker2003laser, glumac2005temporal}. Longenecker \textit{et al} suggested that this agreement results from the existence of separate cooling phases occurring at different periods of time~\cite{longenecker2003laser}.\\
 Mechanisms that could contribute to the reduction in electron temperature (loss of electron energy) include: (1) ambipolar expansion of the plasma followed by adiabatic cooling, (2) elastic and inelastic collisions between free electrons and Ar neutrals, (3) electrons-ions Coulomb collisions, and (4) radiative losses.\\
 Ambipolar expansion was previously shown to be prevalent on the sub-microsecond to microsecond timescale, consistent with $\tau_{T1}$ = 0.41 $\mu$s.
The characteristic time for electron-neutral elastic cooling can be estimated as:

\begin{equation}
    \tau_{e-n} \approx \frac{m_n}{m_e}\frac{1}{\nu_{en}} \approx \frac{m_n}{m_e}\frac{1}{n_n \sigma_{en}v_e}
\end{equation}
Where $n_n$, $m_n$, $\nu_{en}$, $\sigma_{en}$, $m_e$ are the neutral number density, neutral mass, electron-neutral collision frequency, electron-neutral cross-section for momentum transfer, and electron mass, respectively. For argon at 1 atm, $n_n$ $\approx$ 2.5 $\times$ 10$^{19}$ cm$^{-3}$. $\sigma_{en} \approx 10^{-16}$ cm$^2$. For $T_e$ = 5 eV, $v_e \approx 1.3 \times 10^8$ cm/s. We therefore obtain $\tau_{e-n} \approx$ 10-50 $\mu$s. This timescale is consistent with the 24 $\mu$s slower timescale of electron temperature decay.\\
Although electron-ion Coulomb collisions occur at the picosecond timescale, it can be shown that their net effect is to thermalize electrons with themselves, redistributing electron energy rather than removing electron energy. This is a result of the tiny momentum change during the often long range Coulomb collisions between electrons and ions. Similarly, electron energy losses through inelastic electron-neutral collisions and radiation can be neglected compared to the aforementioned loss channels. In summary, our experimental results indicate that electron cooling processes are dominated by ambipolar expansion early in time, followed by electron-neutrals elastic collisions at later times.\\
It is important to note that care was taken to ensure that electron temperatures inferred from the LTS diagnostic were not affected by potential probe heating effects. This was verified by ensuring that similar electron temperatures were obtained when the probe beam pulse energy at 532 nm was increased from 24 mJ to about 30 mJ. Furthermore, the reported error bars mainly account for uncertainties inferred from the fitting procedure.

\begin{figure}
    \centering
    \includegraphics[width=0.70\textwidth]{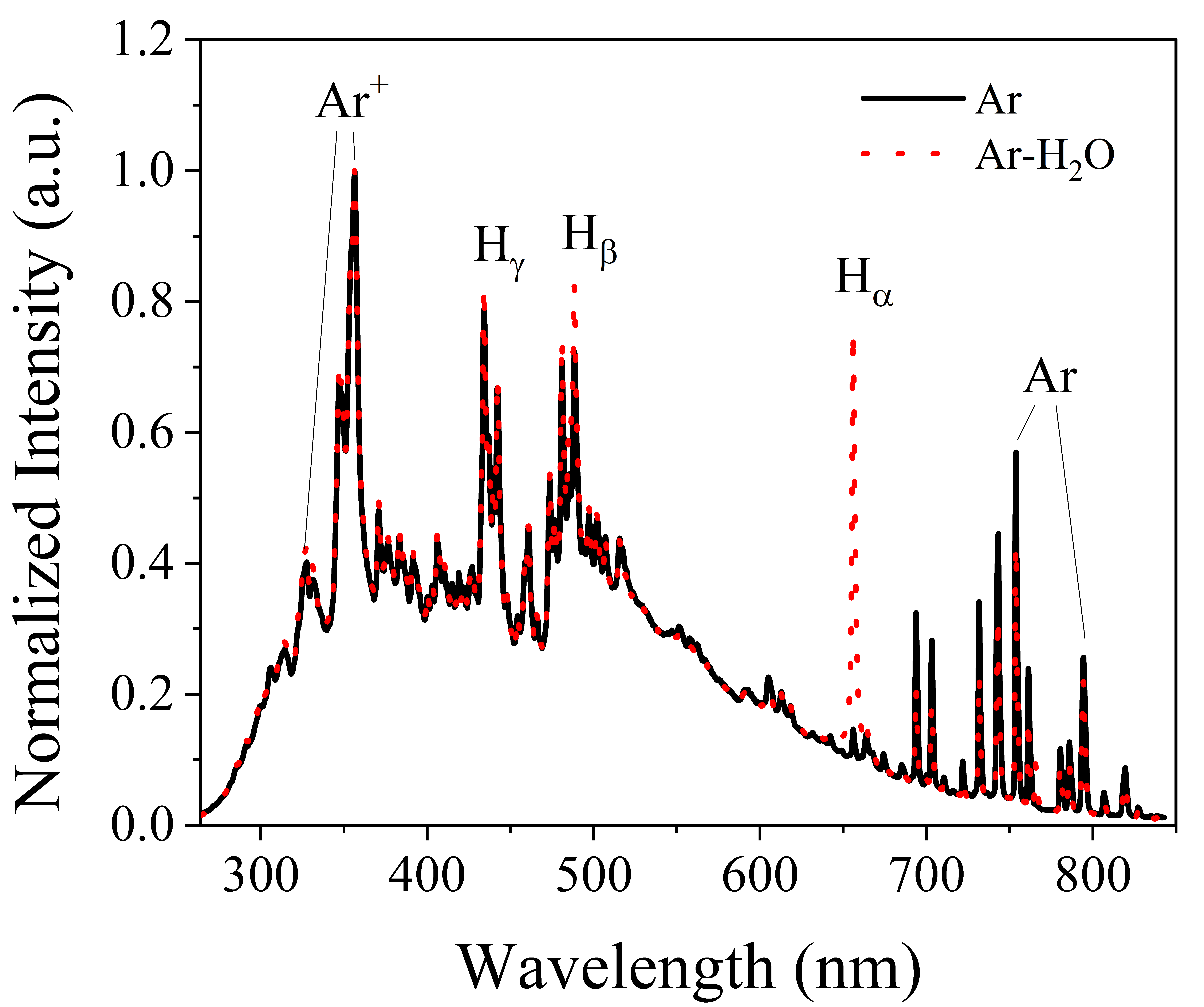}
    \caption{Time-averaged OES spectra measured in Ar and Ar-H$_{2}$O using the miniature pocket spectrometer. H$_\alpha$, H$_\beta$, H$_\gamma$ and a few Ar and Ar$^+$ transitions are visible. The spectral resolution is about 1.5 nm.}
    \label{BroadOES}
\end{figure}

\subsection{Optical Emission Spectroscopy}

OES measurements were deployed to complement LTS experiments, especially given the observed prolonged emission. Before focusing on time-resolved OES-based electron number density measurements, time-averaged, spectrally-broad measurements covering the 200-850 nm spectral range were recorded using a miniature spectrometer that exhibited a spectral resolution \textasciitilde{1.5-2} nm. The spectra were averaged over 260 ms each. As expected, the time-averaged normalized OES spectra depicted in Figure~\ref{BroadOES} are composed of line and continuum radiations. The observed continuum radiation is ascribed to electron-ion and electron-neutral bremsstrahlung processes~\cite{cremers2013handbook, harilal2022optical}. Meanwhile, line radiation features fingerprints of Ar, Ar$^{+}$ and H species among others. Interestingly, it is observed that H atoms are readily present in the pure Ar case. We hypothesized that this observation is related to water vapor impurities in the gas feed lines. Moreover, when comparing normalized spectra for the pure Ar and Ar-H$_{2}$O cases, it appears that they have an almost perfect match in spectral signature, except for the increased intensities of the H atoms Balmer series peaks in the case of Ar-H$_{2}$O. The presence of H atoms in both the pure Ar and Ar-H$_2$O cases will facilitate Stark broadening measurements of the electron number densities. Indeed, without the presence of H atoms, we would have relied on Stark broadening of Ar and Ar$^{+}$ lines. However, the broadening formulas for these species depend on both $n_e$ and $T_e$~\cite{nikiforov2015electron}, which is not the case for the Balmer $\alpha$ and $\beta$ transitions of the H atom in the expected 10$^{16}$-10$^{17}$ cm$^{-3}$ electron number density range.

\begin{figure}
    \centering
    \includegraphics[width=0.95\textwidth]{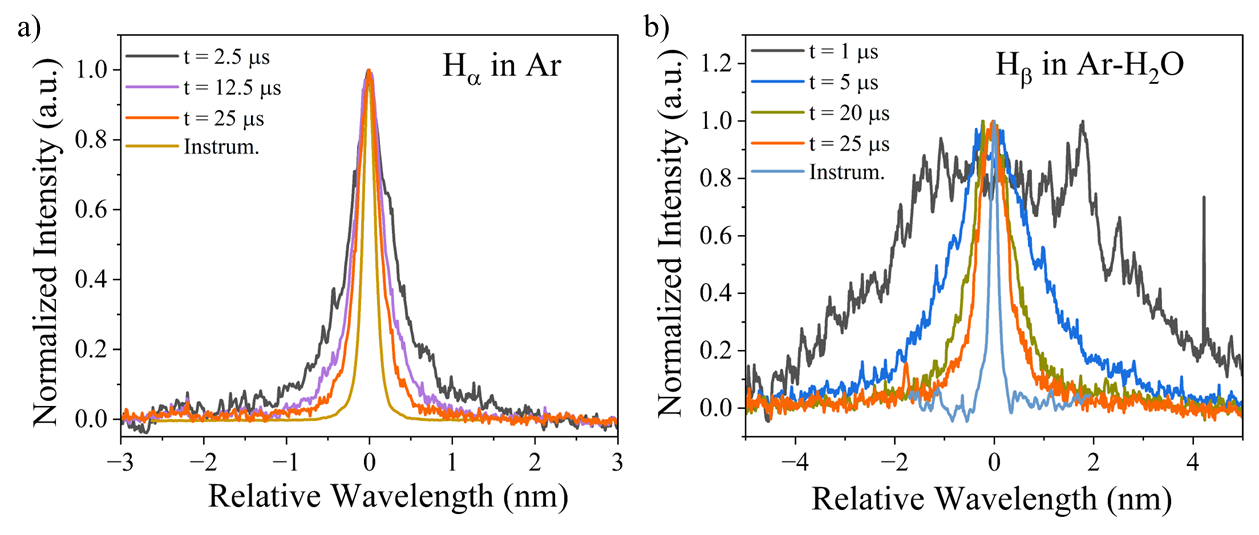}
    \caption{a) Normalized sample H$_{\alpha}$ OES spectra at $t$ = 2.5, 12.5 and 25 $\mu$s in pure Ar. b) Normalized H$_{\beta}$ OES spectra at $t$ = 1, 5, 20 and 25 $\mu$s in Ar-H$_2$O. The camera gate width is 20 ns. Spectrometer entrance slit width = 50 $\mu$m. Instrum. denotes the instrumentation functions measured at 656 and 486 nm, respectively. 15,000 spectra were accumulated for each of the time delay measurements. Spectral resolutions are about 190 and 210 pm at 656 and 486 nm, respectively.}
    \label{OES}
\end{figure}

\begin{figure}[h!]
    \centering
    \includegraphics[width=0.80\textwidth]{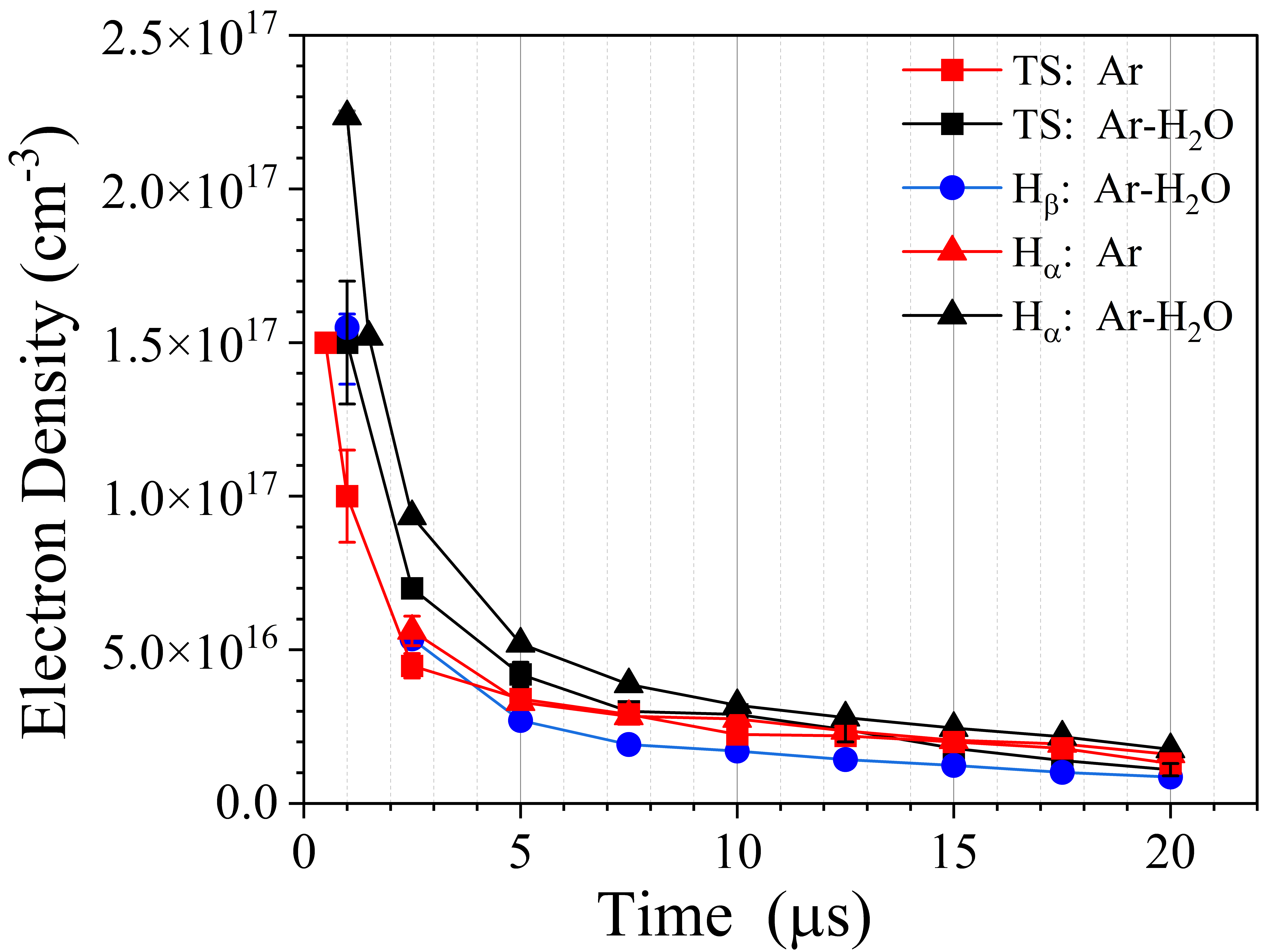}
    \caption{Comparison of time-resolved electron number density inferred from Thomson scattering and OES in Ar and Ar-H$_2$O LPPs.}
    \label{ne}
\end{figure}

\noindent Following these time-averaged measurements, time-resolved, spectrally-broad measurements covering the 500-800 nm spectral range were recorded using the 150 grooves/mm grating of the spectrograph with 20 ns IsCMOS gate widths. Although not shown in this manuscript, from the generation of the laser spark up to about 100 ns after plasma initiation, only broadband continuum emission was observed. After \textit{t} = 120 ns, continuum and line emissions are observed. From these results, it therefore appears that Stark broadening measurements will only be feasible at times later than \textit{t} = 120 ns. Later time measurements will be possible until the width of the broadened OES spectrum becomes comparable to the spectral resolution of the collection system at that wavelength. Interestingly, despite the fact that time-averaged measurements showcased Balmer series $\alpha$ (3 $\rightarrow$ 2), $\beta$ (4 $\rightarrow$ 2) and $\gamma$ (5 $\rightarrow$ 2) transitions in both Ar and Ar-H$_2$O cases, H$_{\beta}$ and H$_{\gamma}$ signal levels in pure Ar were too low when using 20 ns IsCMOS gate widths. Consequently, the time and spectrally-resolved OES measurements will focus on H$_{\alpha}$ in pure Ar, and on H$_{\alpha}$ and H$_{\beta}$ in Ar-H$_{2}$O, respectively.\\
\noindent Temporally and spectrally-resolved OES measurements of the electron number density following Stark broadening of the atomic hydrogen alpha and beta transitions were conducted at the same location as LTS measurements. The same collection system was used for LTS, except that the spectrometer was centered at 656.3 and 486.1 nm, respectively. Figures~\ref{OES}-(a),(b) display sample Balmer alpha and beta spectra, respectively. For the purpose of comparison, the measured instrumental functions at these two wavelengths are overlaid on top of the spectrally-resolved Balmer series spectra. From the latter figures, spectrally-resolved H$_{\alpha}$ and H$_{\beta}$ spectra seem to showcase broadenings comparable to the instrumental functions starting from \textit{t} $\approx$ 25 $\mu$s. Consequently, the OES measurements were only analyzed for the Stark broadening up to \textit{t} = 20 $\mu$s. Note that the sample spectra were collected with the 1200 grooves/mm grating with the overall spectral resolutions of the collection system being 190 pm and 210 pm at 656 and 486 nm, respectively. For each of the time delays, about 15,000 spectra were accumulated (30 seconds exposure time repeated 10 times). Except for H$_{\beta}$ spectra taken early in time (at \textit{t} = 1 $\mu$s), all other OES spectra were fitted using a Voigt profile with a fixed Gaussian contribution corresponding to the measured Gaussian instrumental function.\\
Electron number density estimates were obtained assuming that the Lorentzian component of the Voigt profile was dominated by Stark broadening. Doppler broadening is negligible since upper bound estimates of the H$_\alpha$ and H$_\beta$ FWHMs due to Doppler broadening lead to quantity smaller than spectral resolutions at these two wavelengths. Indeed, assuming a thermal equilibrium between free-electrons and the gas particles, the highest possible gas temperature would be about 7 eV (i.e. about 81,242 K). For the H$_\alpha$ case, such a temperature would correspond to a Doppler FWHM of about 134 pm (for H$_\alpha$, $\Delta \lambda_{D} \approx 4.7 \times 10^{-4} \sqrt{T_g}$, where $T_g$ is the gas temperature in K and $\Delta \lambda_{D}$ is the Doppler broadening FWHM in nm). This value is even lower in the H$_\beta$ case ($\Delta \lambda_{D} \approx 3.5 \times 10^{-4} \sqrt{T_g}$): \textasciitilde{100} pm. If we instead assume a maximum gas temperature of 10,000 K, as measured experimentally by Longenecker et al.~\cite{longenecker2003laser} using Rayleigh scattering under conditions relatively similar to ours (Ar at 1 atm, 20 ns drive beam at 1064 nm but with intensity of 10$^{11}$ W/cm$^{2}$), the figures above change to about 47 and 35 pm for H$_\alpha$ and H$_\beta$, respectively.\\
Van der Waals (VDW) broadening is negligible since pressure relaxation to atmospheric pressure is expected to have occurred after about 1 $\mu$s~\cite{phuoc2005experimental, glumac2005temporal}. The estimated maximum contributions of the VDW broadening mechanism to the measured FWHM of the broadened spectral profiles are therefore about 103 pm and 79 pm for H$_\alpha$ ($\Delta \lambda_{VDW} \approx 5.6 \times P/T_g^{0.7}$, where $P$ is the gas pressure in bar) and H$_\beta$ ($\Delta \lambda_{VDW} \approx 4.3 \times P/T_g^{0.7}$), respectively~\cite{griem2012spectral}. Similarly, one can show that the resonance broadening of these two transitions is also negligible~\cite{simeni2025origins}.\\
\noindent Using the procedure detailed above, the OES-based time-resolved electron number density in Ar and Ar-H$_2$O LPPs are superimposed on the LTS measurements in Figure~\ref{ne}. The reported error bars were inferred from the fitting estimates as well as from background subtraction processing. It can be readily observed that good overall agreement between LTS and Stark broadening measurements was obtained for the entire temporal dynamics spanning 500 ns $\leq$ \textit{t} $\leq$ 20 $\mu$s. The overall agreement is excellent when specifically looking at 10 $\mu$s $\leq$ \textit{ t} $\leq$ 20 $\mu$s. At times earlier than \textit{t} = 10 $\mu$s, all measurement techniques appear to agree within 20\%, except for H$_{\alpha}$ measurements in Ar-H$_2$O, which appear to be consistently higher by about 40-50\%. We attribute the discrepancies between the results of H$_\alpha$ and H$_\beta$ in Ar-H$_{2}$O to a higher continuum background for H$_\beta$ likely with additional corrections required to account for ion dynamics~\cite{gigosos1996new, gigosos2003computer}. We should note that even though free-electrons reach high densities (up to 10$^{17}$ cm$^{-3}$), self-absorption is not expected to be an issue for the OES experiments presented in this manuscript because (1) H atoms constitute a minor species, even in the Ar-H$_2$O case and (2) the absorption path length is short, only \textasciitilde{1-2} mm.

\section{Summary and Conclusions}
We have examined the temporal evolution of nanosecond pulsed Nd:YAG laser-produced plasmas in atmospheric pressure argon and Ar-3\%H$_{2}$O mixtures using broadband imaging, Thomson scattering and optical emission spectroscopy (OES). Time-resolved intensified sCMOS imaging revealed persistent light emission lasting up to 19 ms after spark initiation. Measurements of the free-electron number density obtained from Thomson scattering and OES were in excellent agreement, showing an exponential decay of more than an order of magnitude from \textasciitilde{2} $\times$ 10$^{17}$ cm$^{-3}$ at $t$ = 500 ns to \textasciitilde{1} $\times$ 10$^{16}$ cm$^{-3}$ at $t$ = 20 $\mu$s. This behavior is consistent with processes dominated by ambipolar expansion and two-body radiative recombination at early times while three-body recombination involving Ar$^+$ ions and electrons contributes at later times. Electron temperatures inferred from Thomson scattering measurements over the same time window exhibited a rapid decrease from \textasciitilde{7} eV at $t$ = 500 ns to \textasciitilde{2} eV at $t$ = 2.5 $\mu$s, followed by a more gradual cooling to \textasciitilde{1} eV by $t$ = 20 $\mu$s. Electron density and temperature evolution in pure Ar and in Ar–3\%H$_{2}$O are remarkably similar, indicating that water vapor primarily affects early breakdown and plasma reproducibility, rather than long-term decay dynamics. In terms of limitations of the proposed approach, strong broadband continuum emission at times earlier than 500 ns hindered effective deployment of Thomson scattering. In addition, optical emission spectroscopy after \textasciitilde{20} $\mu$s was limited by the instrument spectral resolution. Future work will extend these measurements to laser-produced plasmas leveraging tin-coated wires as targets for the purpose of investigating the impacts of the plasma parameters on the EUV yield at 13.5 nm. For these experiments, we are planning to deploy a 2-D Thomson scattering setup~\cite{bak2024two, tomita2017time}. Such an approach would facilitate direct comparison of experiments with numerical simulations related to solid~\cite{lezhnin2025particle} and liquid targets~\cite{basko2015structure, basko2016maximum}.

\section{Acknowledgments}
The authors gratefully acknowledge financial support of the U.S. Department of Energy, Office of Fusion Energy Sciences under the Early Career Research award number DE-SC0025627. We gratefully acknowledge the support of the Kavli Foundation Exploration Award in Nanoscience for Sustainability LS-2023-GR-51-2857 and the Carbon Hub. The authors also acknowledge the financial support of the College of Science and Engineering and the Department of Mechanical Engineering of the University of Minnesota, Twin Cities.
\newpage

\section{Appendix A}
\begin{figure}[h!]
    \centering
    \includegraphics[width=0.95\textwidth]{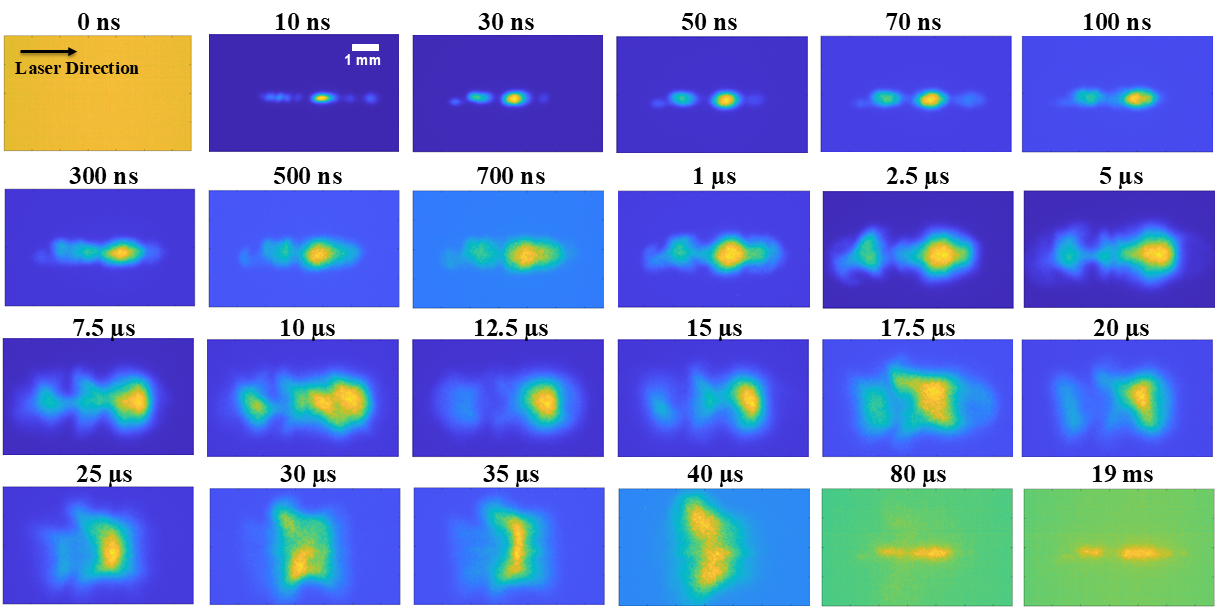}
    \caption{Complete plasma images collage for pure Ar laser sparks. Each image is accumulated over five laser shots. The camera gate width is 5 ns.}
    \label{Pure Ar Collage}
\end{figure}

\begin{figure}[h!]
    \centering
    \includegraphics[width=0.95\textwidth]{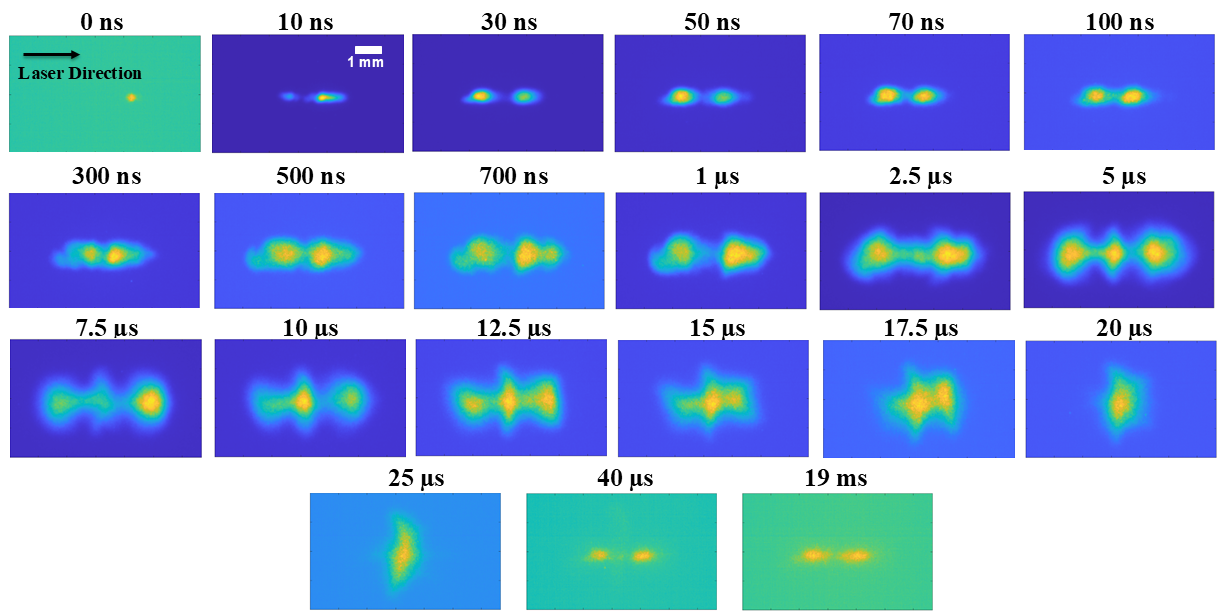}
    \caption{Complete plasma images collage for Ar-H$_{2}$O laser sparks. Each image is accumulated over five laser shots. The camera gate width is 5 ns.}
    \label{Pure Ar-water Collage}
\end{figure}

\newpage
\section{Appendix B}

\begin{figure}[h!]
    \centering
    \includegraphics[width=0.85\textwidth]{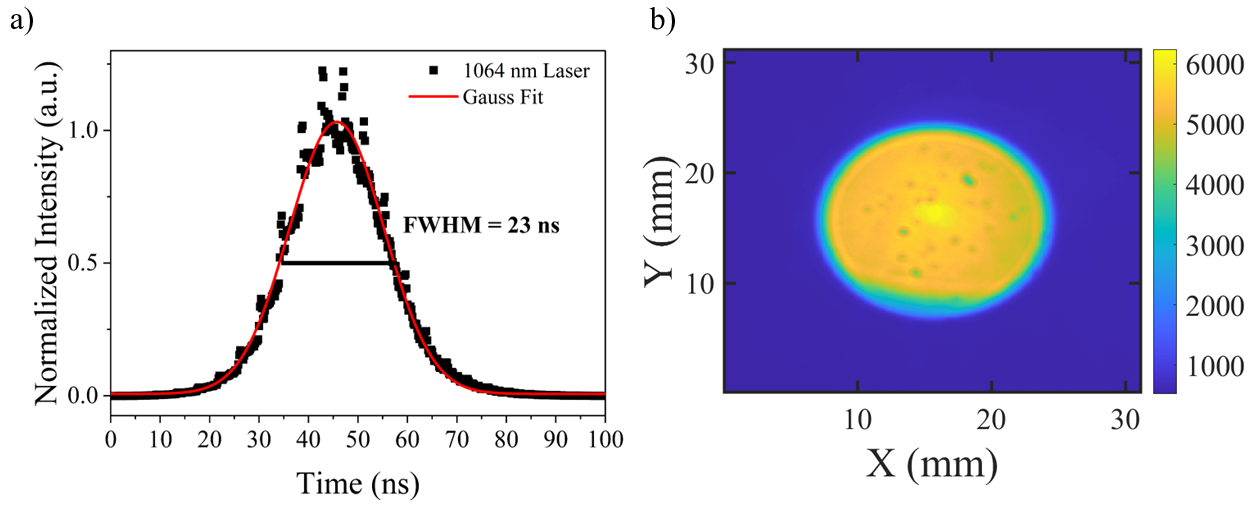}
    \caption{a) 1064 nm drive laser temporal profile. The laser pulse duration is about 23 ns at FWHM. b) Spatial profile of the 1064 nm drive laser beam. The spatial profile shows inhomogeneities and features a flat top.}
    \label{1064 nm profiles}
\end{figure}

\newpage
\bibliographystyle{unsrt} 
\bibliography{Bibio}

\end{document}